\documentstyle[12pt]{article} 
\font\tenrm=cmr10

\def\th{\theta}

\def\De{\Delta}

\def\frac#1#2{{\textstyle{{#1}\over {#2}}}} 
 
\def\lsim{\mathrel{\rlap{\lower4pt\hbox{\hskip1pt$\sim$}} 
    \raise1pt\hbox{$<$}}} 
\def\gsim{\mathrel{\rlap{\lower4pt\hbox{\hskip1pt$\sim$}} 
    \raise1pt\hbox{$>$}}} 
\def\sqr#1#2{{\vcenter{\vbox{\hrule height.#2pt 
         \hbox{\vrule width.#2pt height#1pt \kern#1pt 
         \vrule width.#2pt} 
         \hrule height.#2pt}}}}

\newcommand{\beq}{\begin{equation}} 
\newcommand{\eeq}{\end{equation}} 
\newcommand{\bea}{\begin{eqnarray}}   
\newcommand{\eea}{\end{eqnarray}}  
\newcommand{\rf}[1]{(\ref{#1})}  
  
\renewenvironment{thebibliography}[1]  
 { \rm  
   \begin{list}{\arabic{enumi}.}  
    {\usecounter{enumi} \setlength{\parsep}{0pt}  
     \setlength{\itemsep}{3pt} \settowidth{\labelwidth}{#1.}  
     \sloppy  
    }}{\end{list}}

\newcommand{\ct}[1]{\cite{#1}}

\def\c2th{\cos 2\theta} 
 \def\s2th{\sin2\theta} 
\def\ss{\sin^2 2\theta}
 
\def\secttit#1{\vglue 0.6cm{\bf\large\noindent{#1}}\vglue 0.4cm}

\begin{document}  
\titlepage  
  
\begin{flushright}  
{CCNY--HEP--96/1\\}   
{IUHET 326\\}  
{January 1996\\}  
{hep-ph/9610399 \\}
\end{flushright}  
\vglue 1cm  
	      
\begin{center}   
{ 
{\Large \bf 
Nonequilibrium Neutrino Oscillations  
\\  
in the Early Universe
\\
with an Inverted Neutrino-Mass Hierarchy
\\}  
\vglue 1.0cm  
{V. Alan Kosteleck\'y$^{a}$ and Stuart Samuel$^{b*}$\\}   
\bigskip  
{\it $^a$Physics Department\\}  
%\medskip  
{\it Indiana University\\}  
%\medskip  
{\it Bloomington, IN 47405, U.S.A.\\}  
\vglue 0.3cm  
\bigskip  
 
{\it $^b$Physics Department\\}
%\medskip  
{\it City College of New York\\} 
%\medskip  
{\it New York, NY 10031, U.S.A.\\} 
%\medskip  
  
\vglue 0.8cm  
} 
 
\vglue 0.3cm  
  
{\bf Abstract} 
\end{center}  
{\rightskip=3pc\leftskip=3pc\noindent  
The annihilation of electron-positron pairs 
around one second after the big bang 
distorts the Fermi-Dirac spectrum of neutrino energies. 
We determine the distortions assuming neutrino mixing 
with an inverted neutrino-mass hierarchy.   
Nonequilibrium thermodynamics, 
the Boltzmann equation, 
and numerical integration are used to achieve the results.  
The various types of neutrino behavior
are established as a function of masses and mixing angles.
 
} 
 
\vfill
 
\textwidth 6.5truein
\hrule width 5.cm
\vskip 0.3truecm 
{\tenrm{
\noindent 
$^*$Electronic address: samuel@scisun.sci.ccny.cuny.edu\\}}
 
\newpage  
  
\baselineskip=20pt  
 
%{\bf\large\noindent I.\ Introduction}  
%\vglue 0.2cm   
%\setcounter{section}{1}   
%\setcounter{equation}{0}   

Around a tenth of a second after the big bang,
the early Universe was predominantly a hot lepton-photon gas
in thermal equilibrium
\ct{kt}.
Shortly thereafter,
when the temperature of the Universe fell to about one MeV, 
the electrons and positrons annihilated into photons.  
The transfer of the lepton energy 
to the photon gas made the photon temperature
slightly higher than if
the lepton annihilation had not occurred.  
At about the same time, 
the neutrinos decoupled thermally 
because the frequency of hard-scattering processes 
decreased due to the lower neutrino energies.

Since electron neutrinos couple more strongly 
to electrons and positrons 
than muon or tau neutrinos, 
the electron neutrinos received more energy 
from the electron-positron annihilations.  
In addition, 
higher-energy neutrinos decoupled later 
than lower-energy ones.  
As a result, 
distortions appeared 
in the neutrino Fermi-Dirac distributions.  
The timing of the neutrino decoupling is critical 
for the distortions to arise.  
If the neutrinos had decoupled 
much earlier or much later than 
the electron-positron annihilation, 
then little or no distortion of thermodynamic distributions 
would have arisen.  
One consequence of the timing coincidence 
is that an excess of $\nu_e$ over 
$\nu_\mu$ and $\nu_\tau$ formed around this time.
The distortions have been analyzed 
in the case of no neutrino mixing
in refs.\ \ct{df92a,dt92a}.  

The issue of how the standard model of the early Universe
is affected by the presence of various types of neutrino
and of neutrino masses and mixings
has been investigated by several authors 
during the past few years
\ct{d81a}-\ct{ks95a}.  
In the present work,
we assume that only electron and muon neutrinos
have significant mixing.  
Then,
on average no neutrino oscillations occur 
in the early Universe before the $\nu_e$ excess develops, 
because until then
for every $\nu_e$ that oscillates into a $\nu_\mu$
there is a $\nu_\mu$ that oscillates into a $\nu_e$.  
However,
neutrino mixing can affect 
the generation of the spectral distortions 
\ct{ks95a}.
Also,
after the neutrinos decouple 
and the distortions are established,
energies and densities continue to fall 
as the Universe expands 
and so the relevance of neutrino mixing increases
\ct{kps93a,ks94a}.

The neutrino mixing can be characterized
by the vacuum neutrino-mixing angle $\th$
and the difference $\De = m_2^2 - m_1^2$
between the squared masses of the mass eigenstates.
The parameter region $\De < 0 $ corresponds 
to an inverted neutrino-mass hierarchy, 
in which the heavier mass eigenstate 
has a larger component in the electron neutrino 
than in the muon neutrino.  
For $\Delta < 0 $,
the flavor evolution of neutrinos in the early Universe 
exhibits a rich spectrum of features 
\ct{ks93a}.  
Several recent papers have been based on the possibility 
of an inverted neutrino-mass hierarchy 
\ct{fpq95,cm95,rs96}. 

For $| \Delta | < 10^{-6} $ eV$^2$, 
neutrino mixing effects 
are too small to affect significantly the
production of the $\nu_e$ excess, 
so neutrino oscillations are relevant only after production.
It is therefore a good approximation
to assume the production phase, 
$t < 0.33$ seconds,
generates the known zero-mixing distortions \ct{df92a}
and to adopt the results as input for the oscillation phase
\ct{kps93a,ks93a,ks94a}. 
This procedure avoids
the difficulties of simultaneously 
treating the nonequilibrium neutrino-excess production 
and the neutrino oscillations.

In the region $10^{-6} $ eV$^2 < | \Delta | < 10^{-5} $ eV$^2$, 
oscillations play some role in the distortion 
of neutrino distributions,
while for $| \Delta | > 10^{-5} $ eV$^2$
oscillations definitely must be incorporated 
in the excess production process.  
For $ \Delta  > 10^{-6} $ eV$^2$, 
another approximation is available
to bypass the complexities 
of the full production-oscillation process:
the appropriate initial conditions 
for the pure-oscillation phase can be estimated 
by projecting the zero-mixing neutrino excess onto 
approximate nonlinear mass-eigenstate (ANME) configurations
\ct{ks93a,ks94a}.
It has recently been demonstrated that
ANME configurations are indeed the correct 
initial conditions for the oscillation phase  
for the case of \it positive \rm $\De > 10^{-6}$ eV$^2$
\ct{ks95a}.

To date,
no analysis of neutrino oscillations in the early Universe 
has been performed for 
\it negative \rm $\Delta < - 10^{-6} $ eV$^2$.  
The basic reason is that
early-Universe neutrino oscillations are unstable
when $\De$ is negative,
and so at some point in time the average neutrino flavor
undergoes a dramatic change in behavior.
This time is called the \it instability time \rm 
in what follows.
When $ - 10^{-6} $ eV$^2 < \Delta < 0$, 
the instability time occurs after the production phase, 
and hence does not affect the neutrino excess.  
The method of ANME configurations can therefore be used
\ct{ks93a}.
However,
for $ \Delta < - 10^{-6} $ eV$^2$, 
the instability time occurs during the production phase,
which makes it difficult to surmise appropriate initial conditions 
for the subsequent neutrino oscillations.  

The primary purpose of the present work 
is to analyze this remaining unexplored parameter region.  
Using the full apparatus for the analysis of 
combined oscillations and production,
we have obtained results for the neutrino spectral distortions
and for neutrino oscillations 
for $ \Delta < - 10^{-6} $ eV$^2$. 
A secondary purpose
is to verify the validity of the ANME-based initial conditions 
assumed in ref.\ \ct{ks93a} 
for the oscillation phase
in the region $ - 10^{-6} $ eV$^2 < \Delta < 0$.  

The lengthy equations governing the excess-$\nu_e$ production 
in the presence of oscillations
may be found in section V of ref.\ \ct{ks95a}.  
They are based on nonequilibrium thermodynamics
using the Boltzmann equation,
and they incorporate
hard-scattering production effects from interactions among 
photons, electrons, positrons, and neutrinos,
as well as neutrino forward scattering off neutrinos, 
electrons, and positrons.
Neutrino-neutrino forward scattering 
is often the dominant neutrino-oscillation effect 
by many orders of magnitude.
It renders the system of equations nonlinear,
so numerical simulations are necessary.  
In the present work, 
we employ the notation and methods 
of ref.\ \ct{ks95a}.  
In particular, 
$ v_0 (E,t) 
\equiv ( \nu_e^{\dag}\nu_e + \nu_\mu^{\dag}\nu_\mu )$ 
is the sum of the excess of electron and muon neutrinos 
at energy $E$ per canonical comoving volume per $E/T$, 
$ v_1 (E, t) \equiv 
(\nu_e^{\dag}\nu_e - \nu_\mu^{\dag}\nu_\mu ) $ 
is the difference, 
$ v_4 ( E, t)
\equiv \nu_\tau^{\dag}\nu_\tau $ 
is the excess of tau neutrinos, 
while
$ v_2 ( E, t ) 
\equiv 2 Re ( \nu_e^{\dag} \nu_\mu ) $ 
and 
$ v_3 ( E, t ) 
\equiv 2 Im ( \nu_e^{\dag} \nu_\mu ) $ 
are the real and imaginary parts 
of the off-diagonal component of the density matrix 
for neutrinos with energy $E$.  
In units with $\hbar = c = 1$, 
the canonical comoving volume 
is taken to be $ 1.0 $ MeV$^{-3}$  
when the temperature $T$ of the Universe 
is $1.5$ MeV.  

For positive $ \De$, 
a single behavior characterizes neutrino flavor evolution 
\ct{ks94a}:  
although the flavor may change over time, 
any evolution occurs smoothly.  
In contrast,
when $  - 10^{-6} $ eV$^2 < \Delta < 0$
three types of behavior appear 
\ct{ks93a}   
depending on the values of $\th$ and $\De$. 
Near $\ss = 1$, 
the evolution is smooth.  
For $\ss$ small, 
self-maintained coherence occurs.  
This is a mode exhibiting some features reminiscent of solitons, 
in which a fraction of the neutrinos oscillate essentially in phase. 
It was originally observed in test simulations 
\ct{s93a} 
and is now understood mathematically 
\ct{ks94b}.  
The third behavior, 
called irregular 
in ref.\ \ct{ks93a}, 
occurs for intermediate $\ss$ and
relatively large $ | \Delta |$. 

Mathematical and physical insight 
about irregular behavior has recently been obtained
\ct{s95a}.  
Ignoring the expansion of the Universe 
but incorporating the alignment property 
of the neutrino vectors
\ct{ks93a,ks94a}, 
analytical solutions for the behavior of 
the average neutrino flavor were found.  
These solutions are characterized by two modes.  
When both have very small amplitude, 
smooth behavior occurs.  
When only one mode has sizeable amplitude, 
self-maintained coherence arises.  
Finally,
when both modes are significantly excited, 
bimodal self-maintained coherence appears.  
This can mimic the irregular behavior seen 
in ref.\ \ct{ks93a}.  
It has been conjectured that, 
when the periods of the two modes are incomensurate, 
bimodal self-maintained coherence is irregular behavior
\ct{s95a}.

To verify that incorporating hard-scattering production effects 
leaves unaffected the results of 
ref.\ \ct{ks93a},  
we performed full production-oscillation simulations 
for selected values of $\ss$ and $\Delta$  
in the region $ - 10^{-6} $ eV$^2 < \Delta < 0$.  
These simulations qualitatively confirm the results 
of ref.\ \ct{ks93a} throughout the entire parameter region.  
They also agree quantitatively over all regions displaying 
self-maintained coherence or smooth evolution. 
For example, 
Figures 2 and 4 of ref.\ \ct{ks93a} are reproduced.   
In the irregular region, 
the numerical results are qualitatively the same,
and the quantitative discrepancy can be understood
in terms of bimodal coherence. 
It turns out that the effect of incorporating 
the true neutrino-excess production 
can excite the amplitudes of the two modes 
by slightly different amounts.  
Since the periods are then incommensurate, 
small differences can lead to sizeable numerical changes.  
Bimodal coherence exhibits features similar to chaotic behavior,
in that it is very sensitive to
small changes in parameters and initial conditions.
 
To determine the behavior of a particular simulation, 
one can examine the time evolution 
of the components of average neutrino vector 
$  \left\langle {\vec v} (t) \right\rangle = 
\left\langle \left( { v_1 (t) , v_2 (t) , v_3 (t) } 
\right) \right\rangle $.  
However, 
the single or bimodal nature of coherent oscillations 
is often better displayed as the three-dimensional orbit of 
$\left\langle {\vec v (t) } \right\rangle$.
In the current work, 
bimodal oscillations were evident in 
almost all the simulations in the irregular region.  
Figure 1 shows an example.  
In a few of the simulations, 
the orbits appear irregular:
it is impossible to decide if only two modes occur.  
It may be that some irregular behavior 
is not strictly bimodal coherence 
due to the expansion of the Universe 
or other effects not included in the analysis of 
ref.\ \ct{s95a}.  
 
Figure 2 
shows our results in the $\ss$-$\De$ plane.  
The region labelled `weak bimodal,' 
classified as `residual self-maintained coherence'
in ref.\ \ct{ks93a}, 
is bimodal coherence with one mode 
having a much smaller amplitude
than the other.
It is almost self-maintained coherence 
but exhibits irregular-type behavior 
over sufficiently long time scales.  
Simulation for $\Delta < - 10^{-4} $ eV$^2$ is
prohibitive in computer time, 
but we can determine the behavior
by other arguments.
The essential physics is unchanged from  
$\Delta = - 10^{-4} $ eV$^2$ to $\Delta = - 1 $  eV$^2$,
so irregular behavior should appear in this region.  
For $\Delta < -1$ eV$^2$, 
vacuum oscillation effects dominate 
over the nonlinear forward scattering of neutrinos,
and so vacuum behavior is expected.  
This region is labelled `vacuum decoherence'
in Figure 2.  
The production of excess neutrinos 
quickly decoheres
and neutrinos become vacuum mass eigenstates.  

For positive $\Delta $,
the sum of the excess of electron and muon neutrinos 
is insensitive to the effects of neutrino oscillations 
\ct{ks95a}.
The same is true of the tau-neutrino excess.  
For negative $\Delta$, 
we find similar results.
Hence, our new curves for  
$\left\langle { v_0 }  (t) \right\rangle$ 
and 
$\left\langle { v_4 }  (t) \right\rangle$ 
exhibit little dependence on $\Delta$ and $\ss$.  
They are essentially identical to  
the dashed-dotted and dashed curves in Figure 3 
of ref.\ \ct{ks95a}.  
Likewise, 
the final energy distributions  
for the sum of the excess of electron and muon neutrinos 
and for the tau-neutrino excess 
are insensitive to $\Delta$ and $\ss$ 
and are given 
by the dashed-dotted and dashed curves in Figure 1 
of ref.\ \ct{ks95a}.  

In contrast, 
the remaining three variables
$\left\langle { v_1 } (t) \right\rangle $,  
$\left\langle { v_2 } (t) \right\rangle $,
and
$\left\langle { v_3 } (t) \right\rangle $ 
depend on $\Delta$ and $\ss$.  
The difference
$\left\langle { v_1 } (t) \right\rangle $ 
between the excesses of electron and muon neutrinos 
starts out at zero at $t = 0.08$ seconds 
when $T = 3.0$ MeV,   
rises as excess electron neutrinos are produced, 
and typically declines after the instability time
as many electron neutrinos oscillate into muon neutrinos.  
At the instability time and thereafter, 
$\left\langle { v_2 } (t) \right\rangle $ 
and sometimes 
$\left\langle { v_3 } (t) \right\rangle $ 
become significantly different from zero.  
The sizeable conversion of electron neutrinos into muon neutrinos 
throughout the $\Delta < 0$ region 
is the result of a new resonance conversion mechanism 
\ct{ks93a}
driven by the imaginary part of the off-diagonal component
of the neutrino density matrix, 
which represents the difference between the third components 
of the average vectors for neutrinos and antineutrinos.  

Figure 3 displays 
$\left\langle { v_1 } (t) \right\rangle $,  
$\left\langle { v_2 } (t) \right\rangle $ 
and 
$\left\langle { v_3 } (t) \right\rangle $ 
for the case $\Delta = - 10^{-5}$ eV$^2$ and $\ss = 10^{-3}$, 
a point lying in the irregular region of Figure 2. 
Figure 4 displays 
the same three quantities
for the case $\Delta = - 10^{-6}$ eV$^2$ and $\ss = 10^{-8}$, 
which lies in the weak-bimodal region.  
Rapid fluctuations occur that are not visible 
on the scale of Figure 4. 
In Figure 5, 
we plot the components   
over a short time interval at $t \approx 0.94$ seconds.  
The corresponding orbit is shown in Figure 6.  
Since irregular behavior occurs throughout 
most of the region with $\Delta < -10^{-6}$ eV$^2$
and vectors undergo sizeable unpredictable fluctuations, 
it makes little sense to plot the final production 
profiles for $\vec v$ as a function of energy. 

For $ -10^{-4}$ eV$^2 < \Delta$, 
ANME configurations at early times are flavor eigenstates 
with an excess of electron neutrinos, 
so the first component of $\vec v \left( { E, t } \right) $  
is the largest.  
At later times, 
ANME configurations 
undergo a 180$^o$ rotation in three-space.  
If the nonlinear neutrino-neutrino interactions were absent, 
then little conversion of electron neutrinos 
into muon neutrinos would occur when the rotation is non-adiabatic.  
However, 
after the rotation occurs, 
the neutrino-neutrino interactions render the system unstable 
and a large conversion of electron neutrinos 
into muon neutrinos takes place,
regardless of the rotation rate
\ct{ks93a,ks94a}.
The instability time is expected to occur when 
the vacuum and CP-conserving electron-neutrino terms 
become equal 
\ct{ks93a,ks94a},
i.e., when
\beq
| \Delta | \cos (2 \theta) / ( 2 \bar E ) 
\approx |V_{CP+} | = 
2 \sqrt{2} G_F \bar E \left( { \rho_{e^-} + p_{e^-} + 
\rho_{e^+} + p_{e^+} } \right) / M_W^2 
\quad ,
\label{a}
\eeq
where $\bar E$ is the average neutrino energy,
$G_F$ is Fermi coupling constant, 
$M_W$ is the mass of the $W$ boson, 
and $\rho_{e^\pm}$ and $p_{e^\pm}$ are,
respectively,
the energy densities and pressures 
of the background positrons and electrons.  

When $\theta$ is small so that 
$\cos 2 \theta \approx 1$, 
we find that Eq.\ \rf{a} predicts 
instability times of 
$1.1$, $0.5$, and $0.23$ seconds 
for $\De =-10^{-7}$ eV$^2$, $-10^{-6}$ eV$^2$  
and $-10^{-5}$ eV$^2$,
respectively.  
In our numerical simulations, 
the corresponding instability times are
found to be 
$\sim 1.0$, $\sim 0.45$ and $\sim 0.2$ seconds, 
in good agreement with the theoretical expectations.  

Similarly,
when $\ss$ is near one
we again find good agreement.  
For example, 
when $\ss = 0.49$, 
the instability times occur 
in simulations at 
$\sim 0.55$, $\sim 0.25$ and $\sim 0.12$ seconds 
for $\De = -10^{-6}$ eV$^2$, $-10^{-5}$ eV$^2$  
and $-10^{-4}$ eV$^2$,
respectively.  
The corresponding values from 
Eq.\ \rf{a} are 
$0.52$, $0.24$ and $0.11$ seconds.  

For $\Delta < - 10^{-4}$ eV$^2$, 
the instability time
as determined by Eq.\ \rf{a}
occurs before production begins.  
Hence,
the instability appears
as soon as the electron-neutrino excess 
becomes sufficiently large to render 
the neutrino-neutrino interactions important.  
This happens soon after production begins.  
The instability disrupts the system 
and leads to the irregular behavior seen
in this region in Figure 2.      

Other features of the neutrino oscillations are 
similar to the positive-$\De$ case 
\ct{ks94a}.  
In particular, 
the CP-suppression mechanism is still operative.  
In the regions of coherent oscillations, 
the CP asymmetry of the excess can sometimes 
reach the $1 \%$ level, 
but this is insufficient to affect 
big-bang nucleosynthesis at a level observable
with current technology.  
The evolutions of the antineutrino excesses 
are almost identical to those of the corresponding neutrinos, 
so we have not displayed them.  

To summarize, 
we have obtained the complete phase diagram 
for neutrino oscillations 
for the case of an inverted neutrino-mass hierarchy.  
Five different behaviors are observed:
smooth evolution,
self-maintained coherence, 
bimodal coherence,
irregular behavior, 
and 
vacuum decoherence.
Throughout most of the large-$|\Delta|$ region, 
irregular coherent oscillations are seen. 

%\bigskip 
%\vglue 0.2cm   
\noindent 
\secttit{Acknowledgments} 
 
This work is supported in part  
by the United States Department of Energy 
(grant numbers DE-FG02-91ER40661 and DE-FG02-92ER40698), 
by the Alexander von Humboldt Foundation, 
and by the PSC Board of Higher Education at CUNY. 

\vglue 0.6cm  
{\bf\large\noindent References}  
\vglue 0.4cm  
  
\def\plb #1 #2 #3 {Phys.\ Lett.\ B #1 (19#2) #3.}  
\def\mpl #1 #2 #3 {Mod.\ Phys.\ Lett.\ A #1 (19#2) #3.}  
\def\prl #1 #2 #3 {Phys.\ Rev.\ Lett.\ #1 (19#2) #3.}  
\def\pr #1 #2 #3 {Phys.\ Rev.\ #1 (19#2) #3.}  
\def\prd #1 #2 #3 {Phys.\ Rev.\ D #1 (19#2) #3.}  
\def\npb #1 #2 #3 {Nucl.\ Phys.\ B#1 (19#2) #3.}  
\def\ptp #1 #2 #3 {Prog.\ Theor.\ Phys.\ #1 (19#2) #3.}  
\def\jmp #1 #2 #3 {J.\ Math.\ Phys.\ #1 (19#2) #3.}  
\def\nat #1 #2 #3 {Nature #1 (19#2) #3.}  
\def\prs #1 #2 #3 {Proc.\ Roy.\ Soc.\ (Lon.) A #1 (19#2) #3.}  
\def\ajp #1 #2 #3 {Am.\ J.\ Phys.\ #1 (19#2) #3.}  
\def\lnc #1 #2 #3 {Lett.\ Nuov.\ Cim. #1 (19#2) #3.}  
\def\nc #1 #2 #3 {Nuov.\ Cim.\ A#1 (19#2) #3.}  
\def\jpsj #1 #2 #3 {J.\ Phys.\ Soc.\ Japan #1 (19#2) #3.}  
\def\ant #1 #2 #3 {At. Dat. Nucl. Dat. Tables #1 (19#2) #3.}  
\def\nim #1 #2 #3 {Nucl.\ Instr.\ Meth.\ B#1 (19#2) #3.}

\newpage
\vglue 0.6cm  
{\bf\large\noindent Figure Captions}  
\vglue 0.4cm  

\bigskip
\noindent
Figure 1: An orbit in the irregular region 
for $\ss = 10^{-7}$, $\Delta = - 10^{-5}$eV$^2$ near 
$t = 0.42$ seconds.  
The first mode involves motion around a circle.  
The second mode rotates the circle.  
Only a few periods of the first mode are included.  
When the orbit is displayed over longer time intervals, 
the bimodal coherence produces a figure which 
looks like a ball of wool.  

\bigskip
\noindent
Figure 2: The phase diagram for 
neutrino flavor oscillations 
for an inverted neutrino-mass hierarchy.

\bigskip
\noindent
Figure 3: Time development of the 
nonequilibrium distortions 
for the case $\Delta = -10^{-5}$ eV$^2$ and $\ss = 10^{-3}$. 
(a) The quantity 
$\left\langle { v_1 } (t) \right\rangle$
is shown as a function of time.
(b) The quantity 
$\left\langle { v_2 } (t) \right\rangle$
is shown as a function of time.
Note the change of vertical scale relative to (a).
(c) The quantity 
$\left\langle { v_3 } (t) \right\rangle$
is shown as a function of time.
Note the change of vertical scale relative to (a).

\bigskip
\noindent
Figure 4: Time development of the 
nonequilibrium distortion
for the case $\Delta = -10^{-6}$ eV$^2$ and $\ss = 10^{-8}$. 
(a) The quantity  
$\left\langle { v_1 } (t) \right\rangle$
is shown as a function of time.
(b) The quantity 
$\left\langle { v_2 } (t) \right\rangle$
is shown as a function of time.
(c) The quantity 
$\left\langle { v_3 } (t) \right\rangle$
is shown as a function of time.

\bigskip
\noindent
Figure 5:
Time development of the nonequilibrium distribution 
for the case in Figure 4, 
with $\Delta = -10^{-6}$ eV$^2$ and $\ss = 10^{-8}$. 
The components have been scaled by a factor of about $250$ 
for clarity.  

\bigskip
\noindent
Figure 6: 
Orbit for the case $\Delta = -10^{-6}$ eV$^2$ and $\ss = 10^{-8}$. 
The three quantities displayed in Figure 5 
are combined into a vector and its trajectory is plotted
in three-dimensional space.

\vfill\eject
\end{document}